\documentclass[pre,twocolumn,floatfix,showpacs]{revtex4}
\usepackage{amsmath,amssymb,graphicx,psfrag}

\begin{document}

\title{Ginzburg-Landau theory of crystalline anisotropy for
bcc-liquid interfaces} 

\author{Kuo-An Wu$^1$, Alain Karma$^1$, Jeffrey J. Hoyt$^2$ and  
Mark Asta$^3$}

\address{
$^1$Department of Physics, Northeastern University, Boston, Massachusetts 02115
\break
$^2$Sandia National Laboratories, Albuquerque, New Mexico 87185 \break
$^3$Department of Chemical Engineering and Materials Science, Center for 
Computational Science and Engineering, 
University of California Davis, Davis, CA   95616
}

\begin{abstract}
The weak anisotropy of the interfacial free-energy $\gamma$ is a crucial parameter
influencing dendritic crystal growth morphologies 
in systems with atomically rough solid-liquid interfaces. The 
physical origin and quantitative prediction of this anisotropy are investigated
for body-centered-cubic (bcc)
forming systems using a Ginzburg-Landau theory where the order parameters
are the amplitudes of density waves corresponding to principal reciprocal
lattice vectors. We find that this theory 
predicts the correct sign, $\gamma_{100}>\gamma_{110}$, and magnitude, 
$(\gamma_{100}-\gamma_{110}) / (\gamma_{100}+\gamma_{110})\approx 1\%$, 
of this anisotropy in good agreement with the results    
of MD simulations for Fe. The results show that the 
directional dependence of the rate of 
spatial decay of solid density waves into the liquid,
imposed by the crystal structure, is a main determinant of anisotropy.
This directional dependence is validated by MD computations of density wave
profiles for different reciprocal lattice vectors for $\{110\}$ crystal faces.  
Our results are contrasted with the prediction of the
reverse ordering $\gamma_{100}<\gamma_{110}$ from an earlier formulation of
Ginzburg-Landau theory [Shih \emph{et al.}, Phys. Rev. A {\bf 35}, 2611 (1987)].
 
\end{abstract}

\maketitle

\section{Introduction}

The advent of microscopic 
solvability theory \cite{Langer,KesslerAP,Amar} in the 1980s 
lead to the prediction that the anisotropy of the 
excess free-energy of the crystal-melt interface is 
a crucial parameter that determines the growth rate
and morphology of dendrites, which shape the microstructures
of many commercial metallic alloys. This prediction was
largely validated by phase-field simulations \cite{KarmaRappelII,Provatas} 
of dendritic solidification during the 1990s. 
More recent work in the present decade has focused on the 
quantitative prediction of both the magnitude and the 
anisotropy of the interfacial free-energy $\gamma$
using molecular dynamics (MD) simulations 
\cite{broughton86,davidchack00,davidchack03,davidchack05,hoyt01,hoyt02,hoyt03,ast
a02,morris02,morris03,sun04,sun041,sun05,laird05,hoyt06,song05,davidchack06}. In 
parallel, 
experimental progress has been made to   
determine this anisotropy in metallic systems
from accurate equilibrium shape measurements 
\cite{napolitano,liu01,napolitano04}, which 
extend pioneering measurements of this anisotropy 
in transparent organic crystals \cite{Huang,glicksman}.  

MD-based methods, including the cleaving technique 
\cite{broughton86,davidchack00,davidchack03,davidchack05} and
the capillary fluctuation method (CFM) \cite{hoyt01,morris02}, have been 
successfully 
developed to compute $\gamma$ and to accurately resolve 
its notoriously  small anisotropy of the order of $1\%$. These methods have
been applied to a wide range of systems, including
several elemental face-centered-cubic (fcc) 
\cite{hoyt01,hoyt02,morris02,sun04,sun041},
body-centered-cubic (bcc) \cite{sun04,sun041,hoyt06}, and hexagonal-close-packed
\cite{sun05} metals, as well as
one fcc metallic alloy \cite{asta02} modeled with interatomic potentials 
derived from the embedded-atom-method (EAM), the Lennard-Jones
system \cite{davidchack03,morris03}, hard-sphere 
\cite{davidchack00,song05,davidchack06}
and repulsive power-law potentials \cite{davidchack05}, and, 
most recently, a bcc molecular organic succinonitrile \cite{laird05}
used extensively in experimental studies of crystal growth
patterns.

In systems with an underlying cubic symmetry, the magnitude
of the crystalline anisotropy has been traditionally characterized 
by comparing the values of $\gamma$ corresponding to $\{100\}$ 
and $\{110\}$ crystal faces. MD calculations have yielded anisotropy values   
$(\gamma_{100}-\gamma_{110}) / (\gamma_{100}+\gamma_{110})=0.5-2.5\%$,
and experimental values extracted from equilibrium shape
measurements fall generally within this range. 
What determines physically the positive sign and the magnitude
of this anisotropy, however, remains unclear. One interesting clue
is that MD-calculated anisotropies generally 
depend more on the crystal structure 
than on the microscopic details of inter-molecular forces 
for the same crystal structure, and anisotropies tend to be consistently
smaller for bcc than for fcc \cite{hoyt06}. A striking example  
of the former is the fact that MD studies of Fe \cite{sun04,sun041} and
succninonitrile \cite{laird05}, with the same bcc structure but entirely 
different
inter-molecular forces, have yielded
comparable anisotropies around half a percent.
The weaker anisotropy of $\gamma$ for bcc compared to fcc  
is also consistent with experimental measurements of anisotropy values
in the range of $0.5-0.7\%$ and $2.5-5\%$ 
for the bcc and fcc transparent 
organic crystals succinonitrile and pivalic
acid, respectively \cite{glicksman, muschol}.   

The fact that anisotropy appears to depend more strongly
on crystal structure than inter-molecular
forces suggests that it may be possible to predict this
critical parameter from a continuum 
density wave description 
of the solid-liquid interface, which naturally incorporates
anisotropy because of the broken symmetry of the solid.
The simplest of such descriptions is the   
Ginzburg-Landau (GL) theory of the bcc-liquid interface
developed by Shih \emph{et al.} \cite{Shih}. The 
order parameters of this theory are the amplitudes 
of density waves corresponding to the set of
principal reciprocal lattice vectors $\{\vec K_i\}$, and
the free-energy functional is derived from 
density functional theory (DFT)
\cite{RY,HOI,HOII} with certain 
simplifying assumptions. 

This theory has yielded  
predictions of $\gamma$ for various bcc elements that are
in reasonably good agreement with experiments. Moreover, it has
provided an elegant analytical derivation of the proportionality
between $\gamma$ and the latent heat of fusion, 
in agreement with the scaling relation $\gamma n^{-2/3}=\alpha L$ 
proposed by Turnbull \cite{Turnbull} and recently corroborated
by MD simulations \cite{sun04,davidchack05}, where $n$ is the solid atomic 
density
and $L$ is the latent hear per atom. This theory, 
however, predicts the wrong ordering $\gamma_{100}<\gamma_{110}$.
This apparent failure could be due to
the truncation of larger $|\vec K|$ modes, i.e. density
waves corresponding to reciprocal lattice vectors $\vec K$ with  
$|\vec K|>|\vec K_i|$. 
However, paradoxically, a strong dependence of the anisotropy on larger
$\vec K$ modes would be hard to reconcile with the weak dependence
of this quantity on details of inter-molecular forces seen in
MD studies, since these forces dictate the amplitudes of these
modes in the crystal where the density is sharply peaked 
around atomic positions.

To shed light on this paradox, 
we revisit the simplest GL theory of the 
bcc-liquid interface based on the minimal
set of principal reciprocal lattice vectors. Our 
calculation differs principally from the one of Shih \emph{et al.} \cite{Shih}
in the derivation of the coefficients of the gradient square
terms in the GL free-energy functional. Each term measures the free-energy
cost associated with the spatial variation, in the direction $\hat n$
normal to the solid-liquid interface, of   
a subset of equivalent density waves with the same magnitude
of the direction cosine $\hat K_i \cdot \hat n$ and hence the
same amplitude. Shih \emph{et al.}  
choose these coefficients to be proportional 
to the number of principal reciprocal lattice
vectors in each subset.
This procedure, however, turns out to yield 
an incorrect directional dependence (i.e. dependence on   
$\hat K_i \cdot \hat n$) of the rate of  
decay of density waves into the liquid. Here we find that 
the inclusion of the correct directional dependence, 
as prescribed by DFT, yields the correct 
ordering $\gamma_{100}>\gamma_{110}$
and a reasonable estimate of the magnitude of anisotropy.
Furthermore, we validate this directional dependence 
by MD computations of density wave profiles. 

\section{Ginzburg-Landau theory}

We review the derivation of the
GL theory and compare our results
to MD simulations in the next section.
The theory is derived in the DFT framework
where the free energy of an inhomogeneous system 
is expressed as a functional $F=F[n(\vec{r})]$
of its density distribution $n(\vec{r})$, which can be expanded in the form  
\begin{equation}
n(\vec{r}) = n_0 \left(1+\sum_{i} u_{i}(\vec{r}) \,\, e^{i\vec{K}_{i} \cdot 
\vec{r}}+\dots \right)\label{nrgl}
\end{equation}
where the order parameters $u_{i}$'s are the amplitudes 
of density waves corresponding to 
principle lattice vectors $\langle 110 \rangle$
of the reciprocal fcc lattice, and the contribution
of larger $\vec K$ mode denoted by $``\dots"$ is neglected.  
Expanding the free-energy as a power series of the
$u_{i}$'s around its liquid value $F_l\equiv F[n_0]$ yields 
\begin{eqnarray}
\label{GLF}
\Delta F&=& {n_0 k_B T\over2} \left(\int d\vec{r} \, 
a_2 \sum_{i, j} c_{ij}\,u_{i}\, u_{j}
\,\,\delta_{0,\vec{K}_i+\vec{K}_j}  \right.  \nonumber \\  
&-a_3& \sum_{i,j,k} c_{ijk}\, u_{i} \, u_{j} \, u_{k}\,\, 
\delta_{0,\vec{K}_i+\vec{K}_j+\vec{K}_k} \\ \nonumber
&+a_4& \sum_{i,j,k,l} c_{ijkl} \,u_{i} \, u_{j}\, u_{k}\, u_{l}
\,\,\delta_{0,\vec{K}_i+\vec{K}_j+\vec{K}_k+\vec{K}_l} \\ \nonumber  
&+ b& \left. \sum_{i} c_{i}\,\left|{du_{i}\over dz}\right|^2 \right),
\end{eqnarray}
where we have defined $\Delta F\equiv F-F_l$
and the gradient square terms arise from the spatial variation of
the order parameters along the direction normal to the interface 
parameterized by the coordinate $z$. 
The Kronecker delta
$\delta_{m,n}$, which equals $0$ or $1$ for $m\ne n$ or $m= n$, respectively,
enforces that only combinations of principal reciprocal lattice vectors that form 
closed polygons $\vec{K}_i+\vec{K}_j+\dots=0$ contribute to 
the free-energy functional. 
Closed triangles generate a non-vanishing 
cubic term that makes the bcc-liquid freezing transition first order.
The multiplicative factors   
$a_i$ and $b$ are introduced  
since it is convenient to normalize
the sums of the $c$'s to unity (i.e.  
$\sum_{i} c_{i} = 1$, $\sum_{i,j}c_{ij}\delta_{0,\vec{K}_i+\vec{K}_j}=1$, etc).
To complete the theory, one needs to determine 
all the coefficients appearing in the GL free energy
functional.

The coefficients of the quadratic terms
are obtained from the standard expression for the
free-energy functional that describes 
small density fluctuations of
an inhomogeneous liquid 
\begin{eqnarray}
\Delta F=   
{k_B T \over 2}  \int \int d \vec{r} d \vec r' \delta n(\vec{r}) 
\left[{\delta(\vec r-\vec r') \over n_0}-C(|\vec{r}-\vec r'|)\right] \delta 
n(\vec r')\label{DFT}
\end{eqnarray}
where $\delta n(\vec{r})\equiv n(\vec{r})-n_0 $
and $C(|\vec{r}-\vec r' |)$ is the direct correlation function whose
fourier transform 
\begin{equation}
C(K)=n_0 \int d\vec r\, C(|\vec{r}|) e^{-i \vec{K} \cdot \vec{r}}
\end{equation}
is related to the structure factor  
$S(K)=\left[1-C(K)\right]^{-1}$.

The two expressions for $\Delta F$, Eqs. (\ref{GLF}) and (\ref{DFT}),
can now be related by assuming that the amplitudes 
of density waves vary slowly across the interface on a scale 
$\sim 1/K_{max}$ where $K_{max}$ is the value of $K$ corresponding
to the peak of the structure factor. Accordingly,  
$u_i(z')$ can be expanded in a Taylor series about $z$ 
\begin{eqnarray}
\delta n(\vec r')    
&\approx & n_0 \sum_i 
\left[u_i(z) + {d u_i(z)\over d z} (z'-z)\right. \nonumber \\
& &\left.
+{1\over 2} {d^2 u_i(z)\over d z^2}(z'-z)^2
+\dots \right] e^{i\vec{K}_i \cdot \vec r' }, \label{slow}
\end{eqnarray}
where the contribution $``\dots"$ involving higher-order derivatives can
be neglected. Namely, terms proportional to
$(z'-z)^nd^n u_i(z)/ dz^n \sim 1/(K_{max}w)^n$, where $w$ is the
characteristic width of the solid-liquid interface, i.e. the scale 
over which order parameters vary from a constant value in the solid to zero
in the liquid. Hence, these
terms vanish at large $n$ under the assumption that $w\gg 1/K_{max}$.
Substituting this expression in Eq. (\ref{DFT}) and carrying out
the integral over $\vec r'$, we obtain 
\begin{eqnarray}
\label{DFT2}
\Delta F &\approx& {n_0 k_B T \over 2} \int d \vec{r}
 \left[ \sum_{i,j}{1\over  S(|\vec K_i|)} u_iu_j \delta_{0,\vec{K}_i+\vec{K}_j}
\right. \nonumber \\
& &\left.
-\sum_i {1\over 2} C''(|\vec K_i|)
(\hat{K}_{i} \cdot \hat{n})^2 \left|{du_{i}\over dz}\right|^2 \right],
\end{eqnarray}
where $C''(K)\equiv d^2C(K)/dK^2$.
Comparing Eqs. (\ref{DFT2}) and (\ref{GLF}), we obtain at once
\begin{equation}
a_2\,c_{ij} = {1\over S(|\vec K_{110}|)},\label{a2cii}
\end{equation}
\begin{equation}
b \, c_i = -{1\over 2} C''(|\vec K_{110}|) 
(\hat{K}_{i} \cdot \hat{n})^2, \label{gsq}
\end{equation}
where we have used the fact that
all reciprocal lattice vectors have
the same magnitude $|\vec K_i|=|\vec K_{110}|$.
Summing both sides of Eq. (\ref{a2cii}) and using the normalization
$\sum_{i,j}c_{ij}\delta_{0,\vec{K}_i+\vec{K}_j}=1$ gives
\begin{equation}
a_2 = \sum_{i,j} 
\frac{\delta_{0,\vec{K}_i+\vec{K}_j}}{S(|\vec K_{110}|)} 
= {12\over S(|\vec K_{110}|)},
\label{a2}
\end{equation}
and $c_{ij}=1/12$.
Similarly, summing over $i$
both sides of Eq. (\ref{gsq}), and
using the normalization $\sum_i c_i=1$, yields
\begin{equation} 
b=-{1\over 2} \sum_{i} C''(|\vec K_{110}|) 
(\hat{K}_{i} \cdot \hat{n})^2= - \,2\,C''(|\vec K_{110}|), \label{sum}
\end{equation}
where the second equality can be shown to be independent of
the direction of $\hat n$, and  
\begin{equation} 
c_i={1\over 4}   
(\hat{K}_{i} \cdot \hat{n})^2 \label{ci1}
\end{equation}

To make the difference between the above derivation 
of the $c_i$'s and the one of Shih {\it et al.} explicit, 
consider one of the $\{110\}$ crystal faces  
with $\hat n$ pointing in the $[110]$ direction.
 The set of 12 principal reciprocal 
lattice vectors $\{\vec K_i\}$ corresponding to the  
$\langle 110 \rangle$ directions can be separated into
three subsets with the same value of $(\hat K_i\cdot \hat n)^2$: 
subset I with 8 vectors 
($[011],[0\bar 11],[01\bar 1],[101], 
[\bar 1 01],[10\bar 1],[0\bar 1\bar 1],[\bar 10\bar 1]$) 
and $(\hat K_i\cdot \hat n)^2=1/4$, subset
II with 2 vectors ($[110],[\bar 1\bar 10]$) and 
$(\hat K_i\cdot \hat n)^2=1$, and subset III with 2 vectors  
($\bar 110$, $[1\bar 10]$) and $(\hat K_i\cdot \hat n)^2=0$.   
Density waves
in a given subset have the same amplitude denoted
here by  $u$, $v$, and $w$ for subsets
I, II and III, respectively. 

It follows that the correct coefficient of the
gradient square terms for a given order parameter
$u$, $v$, or $w$, is obtained using the expression for 
$c_i$ given by Eq. (\ref{ci1})  
with the corresponding value of $(\hat K_i\cdot \hat n)^2$
for the corresponding subset I, II, or III, respectively. 
These coefficients are $c_i=1/16$ for
subset I, $c_i=1/4$ for subset II, and $c_i=0$ for subset
III. These coefficients yield the gradient square terms 
$-C''(|\vec K_{110}|)|du/dz|^2$ and
$-C''(|\vec K_{110}|)|dv/dz|^2$ in the GL free energy
functional (\ref{GLF})  since there are $8$ 
equivalent reciprocal lattice vectors in subset I and 
$2$ in subset II,
respectively. The coefficient of $|dw/dz|^2$ vanishes since principal
reciprocal lattice vectors in subset III are orthogonal
to $\hat n$ and $c_i=0$. 

In contrast, Shih {\it et al.} choose the  
$c_i$'s to be equal for all subsets with a 
non-vanishing direction cosine (subsets I and II), 
and $c_i=0$ for subsets with principal
reciprocal lattice vectors orthogonal to $\hat n$ (subset III). 
Since there is a total of 10 reciprocal lattice vectors 
in subsets I and II, the normalization condition 
$\sum_i c_i=1$ yields $c_i=1/10$.
These coefficients yield the gradient square terms 
$-(8/5)C''(|\vec K_{110}|)|du/dz|^2$ and
$-(2/5)C''(|\vec K_{110}|)|dv/dz|^2$ in the GL free energy
functional (\ref{GLF}), which are weighted proportionally
to the number of reciprocal lattice vectors in each subset,
and differ from the correct terms derived above.

For the $\{100\}$ and $\{111\}$ crystal faces, 
the weighting procedure of Shih {\it et al.} 
and Eq. (\ref{ci1}) give coincidentally
the same coefficients of the gradient 
square terms. 
Thus these cases need not be repeated here. The
results for the different crystal faces are
summarized in Table \ref{cis}.

 \begin{table}[b]
 \caption{
Comparison of coefficients of square gradient terms $c_i$
predicted by Eq. (\ref{ci1}) (DFT) and 
Shih \emph{et al.}  \cite{Shih} for the $\{100\}$,
$\{110\}$, and $\{111\}$ crystal faces. For each
orientation, the 12 principal
reciprocal lattice vector are grouped into 
subsets where $\hat K_i\cdot \hat n$ have the same
magnitude in each subset.}
\centering
\begin{tabular*}{0.5\textwidth}%
     {@{\extracolsep{\fill}}ccc|ccc|cc}  \hline
 &
\multicolumn{2}{c}{$100$}&\multicolumn{3}{c}{$110$}&\multicolumn{2}{c}{$111$}
\\ \hline
$(\hat{K}_{i} \cdot \hat{n})^2$ & 0 &1/2& 1/4 & 1 & 0&
0&2/3  \\ \hline
Number of $\vec{K}_{i}$'s    & 4 & 8 & 8
&2 &2 &6& 6 \\ \hline
$c_i$ (Eq. \ref{ci1})  & 0 &1/8&1/16&1/4&0&0&1/6    \\ \hline
$c_i$ (Ref. \cite{Shih}) & 0
& 1/8& 1/10 &1/10& 0 &0 &1/6  \\ \hline
\end{tabular*}
\label{cis}
\end{table}

The determination of all the other coefficients in the
GL free energy functional is identical
to the calculation of Shih {\it et al.}. The coefficients of
the cubic and quartic terms, $c_{ijk}$ and $c_{ijkl}$, respectively,
are determined by the ansatz that all  
polygons with the same number of sides 
have the same weight, which yields $c_{ijk}=1/8$ and
$c_{ijkl}=1/27$; for quadratic terms, this ansatz reproduces
the result $c_{ij}=1/12$ derived above since there are twelve two-sided 
polygons formed by the principle reciprocal lattice vectors.
Using these coefficients and 
identifying each $u_i$ with 
the order parameter 
$u$, $v$, or $w$, depending on whether the 
corresponding ${\vec K_i}$ on one side of a polygon belongs 
to subset I, II, or III, respectively,
Eq. (\ref{GLF}) reduces for $\{110\}$ crystal faces to 
\begin{eqnarray}
\label{eq:energy}
{\Delta F} &=& {n_0 k_B T \over 2} \int d\vec{r} \left[ 
a_2\left({2\over 3} u^2 + {1\over 6} v^2 + {1\over 6} w^2 
\right)\right. \nonumber \\
&-a_3&\left({1\over 2} u^2 v + {1\over 2} u^2 w\right) 
+a_4\left({12\over 27} u^4 + {1\over 27} v^4 +{1\over 27} w^4\right. \nonumber \\
& &\left. +{4\over 27} u^2 v^2
+{4\over 27} u^2 w^2 +{1\over 27} w^2 v^2 +{4\over 27} u^2 v w\right) \nonumber 
\\
& &\left. -C''(|\vec K_{110}|) \left|{du\over dz}\right|^2
-C''(|\vec K_{110}|)\left|{dv\over dz}\right|^2\right],
\end{eqnarray}
The corresponding expression of Shih {\it et al.}
differs by the coefficients of $|du/dz|^2$ and $|dv/dz|^2$
that have an extra multiplicative factor 
of $8/5$ and $2/5$, respectively, as discussed above. 
Their expressions for $\Delta F$ for the 
$\{100\}$ and $\{111\}$ crystal faces are identical
to ours since Eq. (\ref{ci1}) and the equal weight ansatz
yield coincidentally the same $c_i$'s for these faces.
 
Finally, the coefficients $a_3$ and $a_4$ are determined 
by the constraints that the equilibrium state of
the solid is a minimum of free energy, 
$\partial \Delta F / \partial u_i|_{u_i=u_s}$ = 0,
where $u_s$ is the value of all the order parameters in the solid,
and that solid and liquid have equal free energy
at the melting point, $\Delta F(u_s) = 0$. These two
constraints yield the relations $a_3=2a_2/u_s$ and
$a_4 = a_2/u_s^2$ that determine $a_3$ and $a_4$ in terms of
$a_2$ given by Eq. (\ref{a2}), which completes the
determination of all the coefficients.

\section{Results and comparison with MD simulations}

The order parameter profiles 
for $\{110\}$ crystal faces were 
calculated by minimizing $\Delta F$ given
by Eq. (\ref{eq:energy}) with respect
to the order parameters $u$, $v$ and $w$, 
and by solving numerically the resulting set
of coupled ordinary differential equations with
the boundary condition $u=v=w=u_s$ in solid
and $u=v=w=0$ in liquid. The value of $\gamma_{110}$ was computed using
Eq. (\ref{eq:energy}) with these profiles. 
The same procedure was repeated for the $\{100\}$ 
and $\{111\}$ crystal faces. 

We used input parameters for the GL theory computed 
directly from the MD simulations in order to make 
the comparison with these simulations
as quantitative and precise as possible.
These parameters include the peak of the liquid  
structure factor $\approx S(|\vec K_{110}|)$, which yields  
$a_2 = 3.99$ using Eq. (\ref{a2}), 
$C''(|\vec K_{110}|)=-10.40 \,\,{\AA}^2$, and the amplitude of 
density waves corresponding to principal reciprocal
lattice vectors in the solid $u_s=0.72$.

\begin{figure}
\begin{minipage}[h]{8cm}
\includegraphics[width=0.9\textwidth, angle=0]{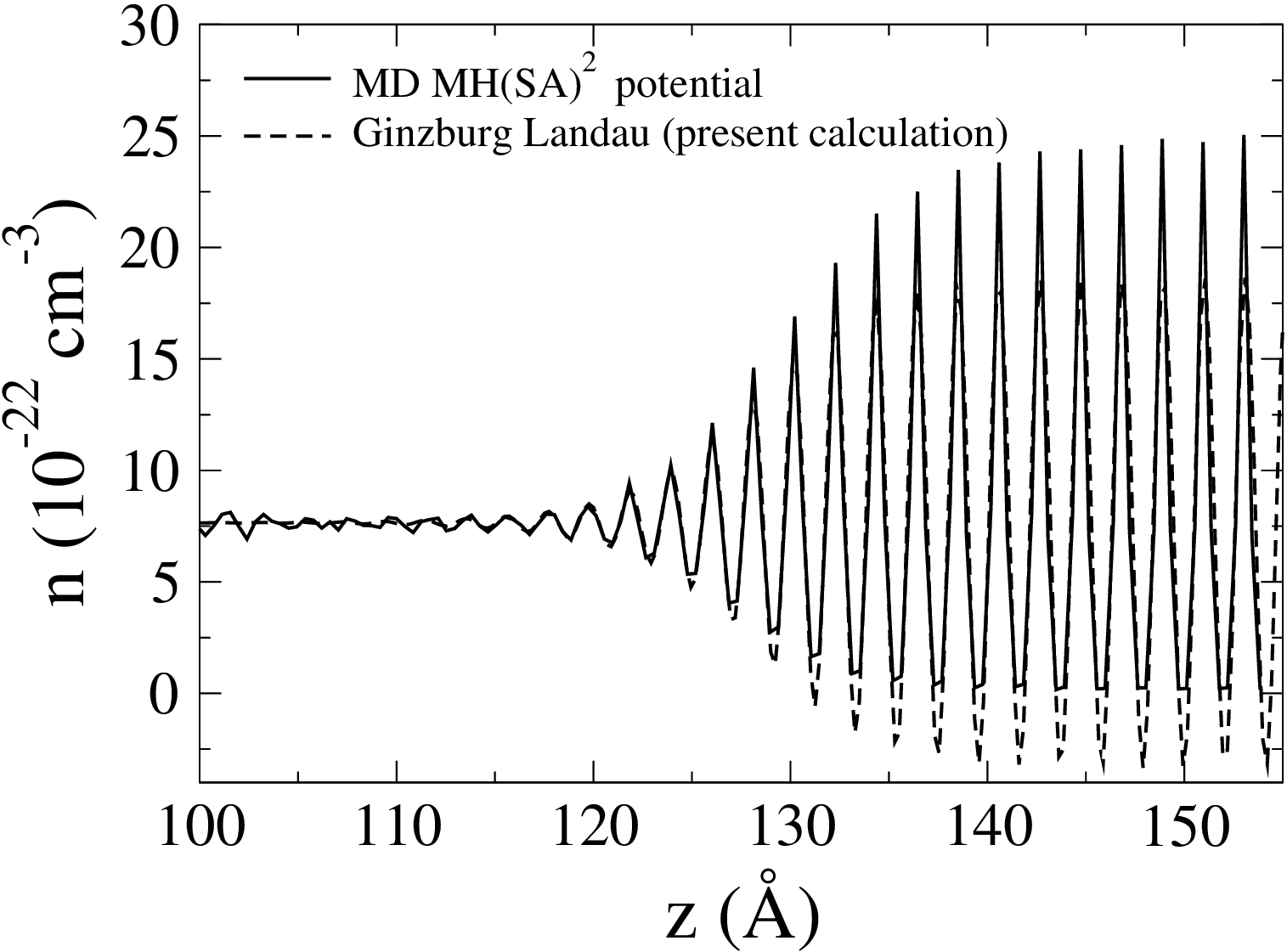} 
\end{minipage}
\caption{Comparison of planar density profiles $n(z)$ from MD
simulations (solid line) and the present
Ginzburg-Landau theory (dashed line) 
for $\{110\}$ crystal faces.} 
\label{nzplot}
\end{figure}

\begin{figure}[b]
\begin{minipage}[t]{8cm}
\includegraphics[width=0.9\textwidth]{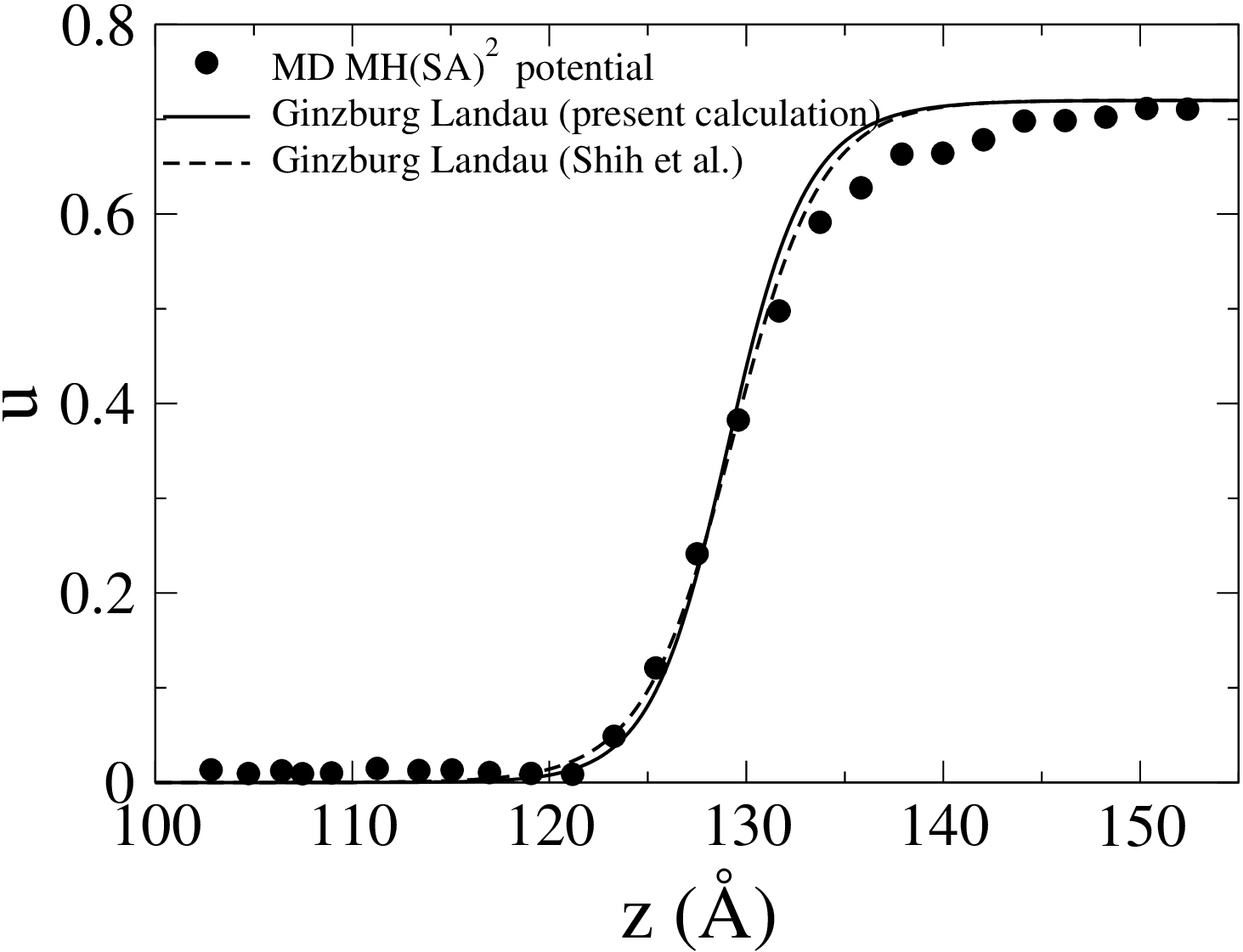} 
\end{minipage}
\hfill
\vskip 1cm
\begin{minipage}[t]{8cm}
\includegraphics[width=0.9\textwidth]{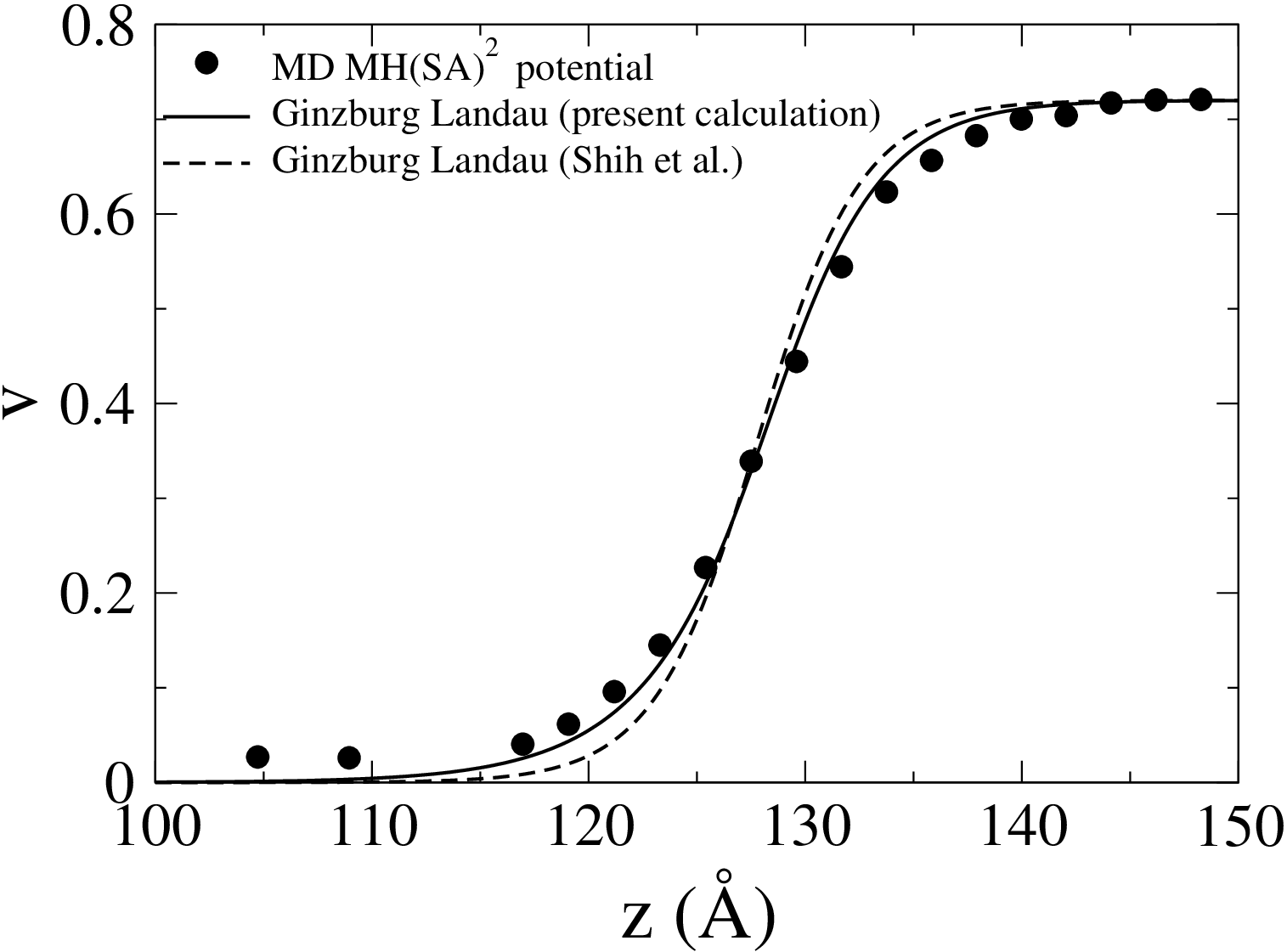} 
\end{minipage}
\hfill
\caption{Comparison of numerically calculated
nonlinear order parameter profiles $u$ and
$v$ for $(110)$ crystal faces obtained from the present
GL theory (solid line) and the GL theory of Shih {\it et al.} \cite{Shih} 
(dashed line) and computed form MD simulations  
using Eq. (\ref{u_i}) with 
$\vec K_{101}$ and $\vec K_{110}$ for
$u$ and $v$, respectively (solid circles).
}
\label{uandvfull}
\end{figure}

\begin{table*}[b]
\caption{Comparison of interfacial free energies for different
crystal faces (in erg/cm$^2$) and anisotropy parameter
$\epsilon_4\equiv (\gamma_{100}-\gamma_{110})/(\gamma_{100}+\gamma_{110})$ 
predicted by Ginzburg-Landau theory with input parameters from MD simulations 
for Fe with the EAM potential of MH(SA)$^2$ \cite{MHSA2} (Table \ref{MDvalues}),  and 
obtained from MD with the MH(SA)$^2$ potential and two other potentials.} 
\centering
\begin{tabular*}{0.75\textwidth}%
     {@{\extracolsep{\fill}}ccccc} \hline
 & $100$ & $110$ & $111$ &$\epsilon_4 (\%)$\\ \hline
MD (ABCH) (Ref. \cite{sun04})& 207.3 (10.1) & 205.7 (10.0) & 205.0 (10.0) & 
0.4(0.4) \\ \hline
MD (Pair) (Ref. \cite{sun04})& 222.5 (14.1) & 220.2 (14.0) & 220.8 (14.0) & 
0.5(0.5) \\ \hline
MD (MH(SA)$^2$) (Ref. \cite{sun04}) & 177.0 (10.8) & 173.5 (10.6) & 173.4 (10.6)
& 1.0(0.6) \\ \hline
GL (present calculation) & 144.26 & 141.35  & 137.57& 1.02 \\ \hline
GL (Shih \emph{et al.} \cite{Shih}) & 144.26 & 145.59 & 137.57 & $-$0.46 \\ 
\hline
\end{tabular*} 
\label{tabgammas}
\end{table*}

The MD simulations were carried out using
the EAM potential for Fe from Mendelev, Han, Srolovitz, Ackland, Sun
and Asta (MH(SA)$^2$) \cite{MHSA2} 
and the same thermodynamic ensemble and geometries
as in Ref. \cite{sun041}, which need not be
repeated here. The main difference of
the present simulations is the way in which the MD results were used 
to calculate density wave profiles. 
In Ref. \cite{sun041}, the amplitudes were computed by averaging
over many configurations the instantaneous value of a planar
structure function (i.e., the magnitude of the complex Fourier 
coefficients of the density).  With this approach the amplitudes
of density waves saturate to a small
non-vanishing value in the liquid. These amplitudes,
however, are generally expected to vanish in the liquid that
has no long range order, consistent with the GL theory.

To calculate amplitudes that vanish in the liquid, the
following procedure is followed.   
We first compute the average number density   
$n(\vec r)=n(x,y,z)$, with $z$ measured
from a fixed reference plane of atoms in the solid. 
During the MD simulations, only those  
configurations where
the solid-liquid interface has the same average position
along $z$ were considered.  As described in detail by Davidchack
and Laird, \cite{Davidchack98}
this procedure avoids an artifical broadening of the
density profiles due to either the natural fluctuations in the average 
position of the interface or Brownian motion of the crystal.
The interface position is found by first assigning to each 
atom an order parameter 
proportional to the mean square displacement
of atoms from their positions on a perfect bcc lattice.  
(The order parameter calculation is the same as that used in the 
capillary fluctuation method and is described in more detail in
reference \cite{sun04}).  Then an order
parameter profile as a function of $z$ is computed by averaging within
the $x-y$ plane and the interface position is that value of $z$ where 
the averaged order parameter is midway between the bulk liquid and bulk 
solid values. 

Next, we compute the $x-y$ averaged density
\begin{equation}
n(z)=\frac{1}{L_xL_y}\int_0^{L_x} \int_0^{L_y} 
dxdy \, n(\vec r),\label{nz}
\end{equation}
which is illustrated in Fig. \ref{nzplot}.
Lastly, we calculate the
amplitude of density waves  
from the fourier transform
\begin{equation}
u_i=\frac{1}{L_xL_y \Delta z}
\int_0^{L_x} \int_0^{L_y}\int_{z_j}^{z_{j+1}} dx dy dz \, n(\vec r) 
\exp(i\vec K_i \cdot \vec r),\label{u_i}
\end{equation} 
where $z_j$ and $z_{j+1}$ correspond
to sequential minima of $n(z)$ 
and $\Delta z \equiv z_{j+1}-z_j$. In addition,
$u_i$ is evaluated at the midpoint 
of this interval, $(z_j+z_{j+1})/2$. The order
parameters $u$ and $v$ were computed for $\{110\}$
crystal faces using $\vec K_{110}$ and $\vec K_{101}$,
respectively.

The results of the present GL theory are compared to
those of Shih \emph{et al.} and MD simulations
in Fig. \ref{uandvfull} and Table \ref{tabgammas}.
Using Eq. (\ref{eq:energy}) with the $c_i$'s
given by Eq. (\ref{ci1}), we obtain the correct 
ordering of interfacial free energies $\gamma_{100} > \gamma_{110}$ 
and a weak capillary anisotropy $(\gamma_{100}-\gamma_{110})/ 
(\gamma_{100}+\gamma_{110})\approx 1\%$,   
consistent with the results of MD simulations for bcc elements 
\cite{davidchack05,sun04,sun041,laird05,hoyt06}, while the 
ansatz of equally weighted $c_i$'s of Shih {\it et al.} (with the
values listed in Table \ref{cis}) 
gives the reversed ordering $\gamma_{100} < \gamma_{110} $.
Note that the predictions of GL theory are to be compared
to the MD results with the MH(SA)$^2$ potential in Table \ref{tabgammas} 
since this potential is used here to 
compute input parameters for this theory given in
Table \ref{MDvalues}. MD results for the other  
potentials are mainly included to illustrate the dependence of
$\gamma$ and its anisotropy on details of interatomic forces.
  
Fig. \ref{nzplot} shows that
the planar density profile predicted by  
GL theory, obtained by substituting Eq. (\ref{nrgl}) with
the numerically calculated order parameter profiles into
Eq. (\ref{nz}), is in remarkably good agreement with MD simulations  
on the liquid side. The discrepancy on the solid side is due to the
fact that GL theory neglects the contribution of larger
$|\vec K|$ reciprocal lattice vectors that contribute to the
localization of density peaks around bcc lattice
positions in MD simulation.

\begin{figure} [t]
\begin{minipage}[h]{8cm}
\includegraphics[width=0.9\textwidth]{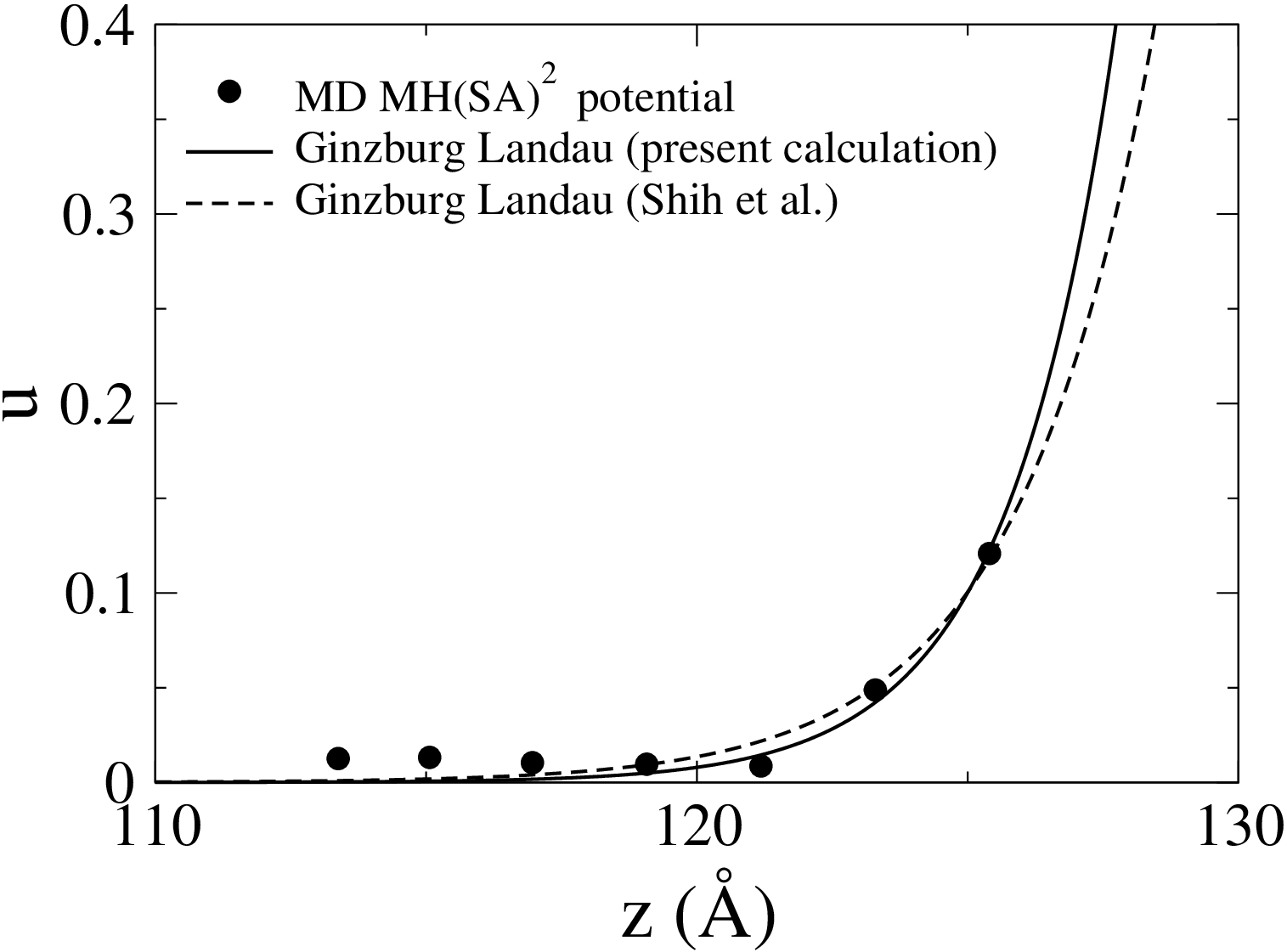} 
\end{minipage}
\vskip 1cm
\hfill
\begin{minipage}[h]{8cm}
\includegraphics[width=0.9\textwidth]{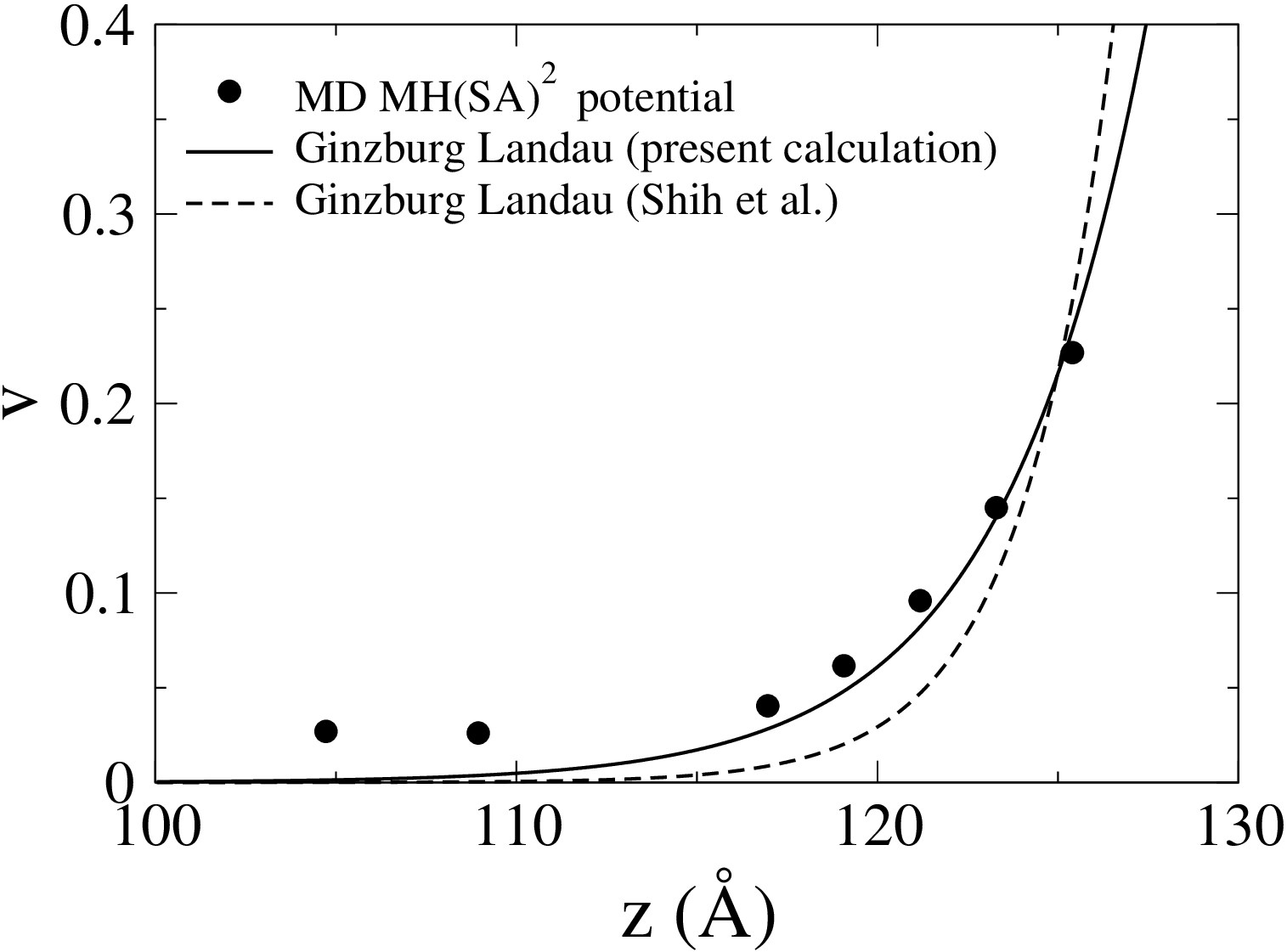} 
\end{minipage}
\hfill
\caption{ 
Comparison of analytically calculated linearized
order parameter profiles $u$ and
$v $ for $(110)$ crystal faces near the liquid from the present 
GL theory (solid line) and the
GL theory of Shih {\it et al.} \cite{Shih} 
(dashed line), and computed from MD
simulations using Eq. (\ref{u_i}) 
with $\vec K_{101}$ and $\vec K_{110}$ for
$u$ and $v$, respectively (solid circles).
}
\label{uandvlinear}
\end{figure}

Fig. \ref{uandvfull} shows that the amplitude profiles and the
interface widths predicted by GL theory 
are in good agreement with MD simulations. 
The MD results clearly validate the directional dependence of the rate
of spatial decay of density waves in the liquid that is the main determinant
of the anisotropy of $\gamma$. This directional
dependence is most clearly seen by examining the amplitude
profiles on the liquid side of the interface. In this region,
the amplitudes of density waves are sufficiently small that 
one can neglect the cubic and quartic terms in 
the GL free energy functional. The resulting linear
second order differential equations for $u$ and $v$ obtained
by minimizing this functional 
can be solved analytically, and have exponentially decaying
solutions that are compared to the MD results in Fig. \ref{uandvlinear}.
The coefficients of the gradient square terms in the free energy
functional control the decay rates. The $u$ and $v$ profiles 
(i.e. the amplitude of density
waves corresponding to 
$\vec K_{101}$ and $\vec K_{110}$) calculated
with coefficients that depend on the 
angle between the principal reciprocal lattice
vectors and the interface normal through Eq. (\ref{ci1}),
which is consistent with DFT, have different decay
rates that are in good quantitative 
agreement with the MD results. In contrast, $u$ and $v$
profiles calculated based on the ansatz 
of equal weights for the $c_i$'s \cite{Shih}
have the same spatial decay rate, which does not agree 
with the MD results.

It is interesting to note that Mikheev and Chernov (MC) \cite{MC1,MC2},
in a formulation of the anisotropy of the solid-liquid interface 
$mobility$, also stress the importance of the decay rate    
of the amplitude of density waves.
The MC model predicts crystal growth rates and anisotropies that are 
in qualitative agreement with MD simulations of FCC systems.  The theory,
however, is linear in the sense that only the effective widths of the 
density profiles, which are allowed to vary with $\vec K$ and $\hat n$,
are required and the authors make no attempt to compute, as was done here,
the full amplitude profile as a function of $z$.

Finally, even though we have focused primarily in this paper
on crystalline anisotropy, it is useful to re-examine the prediction
of GL theory for the magnitude of $\gamma$ and for the Turnbull coefficient
using input parameters from the present MD simulations.
Shih {\it et al.} \cite{Shih} derived an analytical expression for the magnitude
of $\gamma$ in the isotropic approximation where all the order parameters
are assumed to have the same profile through the interface, i.e. $u_i(z)=u$
for all $i$. In this approximation, the free energy density
reduces to the sum of the gradient square term  
$b|du/dz|^2$ and a quartic polynomial in $u$. The stationary
profile $u(z)$ that minimizes the free energy  
is then an exact hyperbolic tangent profile and the analytical
expression for the interfacial energy is  
\begin{equation}
\gamma = {n_0 k_B T_m \over 6} u_s^2 (a_2 b)^{1/2}. \label{gammadef} 
\end{equation}
Furthermore, Shih {\it et al.} related
the latent heat (per atom) to the temperature variation 
of the inverse of the peak of structure
factor proportional to $a_2$ (Eq. \ref{a2}), 
\begin{equation}
L = {T_m\over N } \left. {\partial \Delta F \over \partial T}
\right|_{T=T_m}  
= {k_B T_m^2 \over 2}u_s^2 \left. \frac{da_2}{dT} \right|_{T=T_m} 
\label{Ldef} 
\end{equation}
where $N$ is the number of atoms in the system. This yields  
the expression for the Turbull coefficient
\begin{equation}
\alpha=\frac{\gamma n_0^{-2/3}}{L}
=\frac{n_0^{1/3}(a_2b)^{1/2}}{3T_m\left. da_2/dT\right|_{T=T_m}},
\label{TurnbullGL}
\end{equation}  
which Shih {\it et al.} evaluate using parameters
for the hard sphere system \cite{Shih}.
Using the values
of the various coefficients obtained from MD simulations listed
in table \ref{MDvalues}, Eq. (\ref{gammadef}) 
yields a value of $\gamma=147.4$ erg/cm$^2$
in reasonably good agreement with the average values of $\gamma$
for the different crystal faces in Table \ref{tabgammas}
obtained from MD simulations and the fully anisotropic
GL calculation with different order parameter profiles. With
the same coefficients, Eq. (\ref{Ldef}) yields a latent heat value
$L=0.114$ eV/atom about 30\% lower than the MD value 
in Table \ref{MDvalues}, where the difference can be attributed
to the contribution of larger $\vec K$ modes that are 
neglected in GL theory. Eq. (\ref{TurnbullGL}) in turn predicts
a value of the Turnbull coefficient $\alpha = 0.45$ that
is about 25\% larger than the MD value $\alpha \approx 0.36$,
owing to the underestimation of the latent heat of melting
in GL theory with input parameters of Table \ref{MDvalues}
from the present MD simulations. 
In the future, it would be interesting to test how    
the Turbull coefficient predicted by GL theory (Eq. \ref{TurnbullGL})
varies with input parameters computed from MD simulations using 
different interatomic potentials.

\begin{table*}[b]
\caption{Values of input coefficients for Ginzburg-Landau theory 
computed from MD simulations using the EAM potential for Fe
from MH(SA)$^2$ \cite{MHSA2} and average
value of $\gamma$ and latent heat of melting from these simulations.}
\centering
\begin{tabular*}{0.75\textwidth}%
     {@{\extracolsep{\fill}}ccccccc} \hline
 & $a_2$ & $b$ (${\AA}^2$) & $da_2/dT$ (K$^{-1}$) &$u_s$& $\gamma$ 
(erg/cm$^2$)& $L$ (eV/atom) \\ \hline
MD (MH(SA)$^2$) & 3.99 & 20.81 & 0.00163 & 0.72 & 175(11) & 0.162 \\ \hline
\end{tabular*} 
\label{MDvalues}
\end{table*}

\section{Conclusions}

We have revisited the simplest GL theory of the
bcc-liquid interface whose order parameters are the amplitudes
of density waves corresponding to principle reciprocal lattice
vectors. We find that, despite its simplicity, this theory
is able to predict the density wave structure of the interface
and the anisotropy of the interfacial energy, in reasonably
good quantitative agreement with the results of MD simulations.
 
A main determinant of the anisotropy of the interfacial energy
in this theory is the rate of spatial decay 
of density waves in the liquid. This decay rate
must depend on the angle between principal reciprocal
lattice vectors and the direction normal to the interface for this
theory to be consistent with DFT.  This directional dependence, 
which we validated quantitatively by MD simulations,
is a direct reflection of the underlying crystal structure. Therefore,
the present results provide a simple physical picture of the 
strong relationship between crystal structure and crystalline anisotropy,
consistent with the findings of a growing body of MD-based and
experimental studies of crystalline anisotropy.

An interesting future prospect is to extend the GL theory
to other crystal structures, and in particular 
fcc-liquid interfaces. This requires, however, to consider the
coupling of density waves corresponding to
the principal reciprocal lattice vectors to 
larger $\vec K$ modes, which makes the theory intrinsically
more complicated.

\begin{acknowledgments}
This research is supported by U.S. 
DOE through Grants No. DE-FG02-92ER45471 (KW and AK) and
No. DE-FG02-01ER45910 (JJH and MA)
as well as the DOE Computational 
Materials Science Network program. 
Sandia is a multiprogram laboratory operated by Sandia Corporation, a
Lockheed Martin Company, for the DOE's National Nuclear Security
Administration under contract DE-AC04-94AL85000.

\end{acknowledgments}


\begin{thebibliography}{99}

\bibitem{Langer} J. S. Langer, in {\it Chance and Matter}, Lectures on the 
Theory of Pattern Formation, Les Houches, Session XLVI, edited by J. Souletie, 
J. Vannimenus, and R. Stora (North-Holland, Amsterdam, 1987), pp. 629-711.

\bibitem{KesslerAP} D. Kessler, J. Koplik, and H. Levine, Adv. Phys. {\bf 37}, 
255 (1988).


\bibitem{Amar} M. Ben Amar and E. Brener, Phys. Rev. Lett. {\bf 71}, 589 (1993).

\bibitem{KarmaRappelII} A. Karma and W. J. Rappel, Phys. Rev. Lett. {\bf 77}, 
4050 (1996); Phys. Rev. E {\bf 57}, 4323 (1998).
\bibitem{Provatas} N. Provatas, N. Goldenfeld, and J. Dantzig, Phys. Rev. Lett.
{\bf 80}, 3308 (1998).

\bibitem{broughton86}
J. Q. Broughton and G. H. Gilmer, J. Chem. Phys. {\bf 84}, 5759 (1986).

\bibitem{davidchack00}
R. L. Davidchack and B. B. Laird, Phys. Rev. Lett. {\bf 85}, 4751 (2000).

\bibitem{davidchack03}
R. L. Davidchack and B. B. Laird, J. Chem. Phys. {\bf 118}, 7651 (2003).

\bibitem{davidchack05}
R. L. Davidchack and B. B. Laird, Phys. Rev. Lett. {\bf 94}, 086102 (2005).

\bibitem{hoyt01}
J. J. Hoyt, M. Asta and A. Karma, Phys. Rev. Lett. {\bf 86}, 5530 (2001).

\bibitem{hoyt02}
J. J. Hoyt and M. Asta, Phys. Rev. B {\bf 65}, 214106 (2002).

\bibitem{hoyt03}
J. J. Hoyt, M. Asta and A. Karma, Mat. Sci. Engin. R {\bf 41}, 121 (2003).

\bibitem{asta02}
M. Asta, J. J. Hoyt and A. Karma, Phys. Rev. B {\bf 66}, 100101(R) (2002).

\bibitem{morris02}
J. R. Morris, Phys. Rev. B {\bf 66}, 144104 (2002).

\bibitem{morris03}
J. R. Morris and X. Y. Song, J. Chem. Phys. {\bf 119}, 3920 (2003).

\bibitem{sun04}
D. Y. Sun, M. Asta, J. J. Hoyt, M. I. Mendelev and D. J. Srolovitz, Phys.
Rev. B {\bf 69}, 020102(R) (2004).

\bibitem{sun041}
D. Y. Sun, M. Asta and J. J. Hoyt, Phys.
Rev. B {\bf 69}, 174103 (2004).

\bibitem{sun05}
D. Y. Sun, M. I. Mendelev, C. Becker, M. Asta, K. Kudin, D. J. Srolovitz, J. J.
Hoyt, T. Haxhimali and A. Karma, Phys. Rev. B (submitted).

\bibitem{laird05}
X. Feng and B. B. Laird, J. Chem. Phys. (submitted).

\bibitem{hoyt06}
J. J. Hoyt, M. Asta and D. Y. Sun, Phil. Mag. (in press).

\bibitem{song05}
Y. Mu, A. Houk and x. Y. Song, J. Phys. Chem. B {\bf 109}, 6500 (2005).

\bibitem{davidchack06}
R. L. Davidchack, J. R. Morris and B. B. Laird, J. Chem. Phys. (submitted).

\bibitem{napolitano}
R. E. Napolitano, S. Liu and R. Trivedi, Interf. Sci. {\bf 10}, 217 (2002).

\bibitem{liu01}
S. Liu, R. E. Napolitano and E. Trivedi, Acta Mater. {\bf 49} 4271 (2001).

\bibitem{napolitano04}
R. E. Napolitano and S. Liu, Phys. Rev. B {\bf 70}, 214103 (2004).

\bibitem{Huang} S.-C. Huang and M. E. Glicksman, Acta Metall \textbf{29}, 701 
(1981).

\bibitem{glicksman} M. E. Glicksman and N. B. Singh, J. Cryst. Growth {\bf 98}, 
277 (1989).

\bibitem{muschol} M. Muschol, D. Liu and H. Z. Cummins, Phys. Rev. A {\bf 46}, 
1038 (1992).

\bibitem{Shih} W. H. Shih, Z. Q. Wang, X. C. Zeng and D. Stroud, Phys.
Rev. A {\bf 35}, 2611 (1987).

\bibitem{RY} T. V. Ramakrishnan and M. Yussouff, Phys. Rev. B 
 {\bf 19}, 2775 (1979).

\bibitem{HOI} A. D. J. Haymet and D. Oxtoby, J. Chem. Phys. 
 {\bf 74}, 2559 (1981).

\bibitem{HOII} D. W. Oxtoby and A. D. J. Haymet, J. Chem. Phys. 
 {\bf 76}, 6262 (1982).

\bibitem{Turnbull} D. Turnbull, J. Appl. Phys. {\bf 24}, 1022 (1950).





\bibitem{MHSA2} M. I. Mendelev, S. Han, D. J. Srolovitz,
  G. J. Ackland, D. Y. Sun and M. Asta, Philos. Mag. {\bf 83}, 3977 (2003).
\bibitem{Davidchack98}
R. L. Davidchack and B. B. Laird, J. Chem. Phys., {\bf 108}, 9452 (1998).

\bibitem{MC1} L. V. Mikheev and A. A. Chernov, Sov. Phys. JETP
 {\bf 65}, 971 (1987).

\bibitem{MC2} L. V. Mikheev and A. A. Chernov, J. Cryst. Growth
 {\bf 112}, 591 (1991).

\end{thebibliography}
\end{document}